\documentclass[a4paper,11pt]{article}
\usepackage{pos}
\usepackage{amsmath}
\usepackage{braket}
\usepackage{tikz-feynman}

\title{Reconstruction of the vector meson propagator using a generalized eigenvalue problem}
\author*[a]{Fabian J. Frech}
\author[b]{Finn M. Stokes}
\author[c]{Kalman K. Szabo}
 \onbehalf{for the Budapest-Marseille-Wuppertal collaboration}

\affiliation[a]{Bergische Universität Wuppertal, D-42119 Wuppertal, Germany}
\affiliation[b]{Special Research Centre for the Subatomic Structure of Matter, Department of Physics University of Adelaide, South Australia 5005, Australia}
\affiliation[c]{J\"ulich Supercomputing Centre, Forschungszentrum J\"ulich, D-52425 J\"ulich, Germany}

\emailAdd{frech@uni-wuppertal.de}

\abstract{

For long distances in the euclidean time the vector-vector correlator ($\rho$) has an exponentially decreasing signal-to-noise ratio.
However, the vector correlator not only consists of the vector meson but also receives contributions from a two-pion system with the same quantum numbers. We measure all two-pion propagators with an energy lower than the mass of the resting vector meson and employ a generalized eigenvalue problem (GEVP) to resolve the different contributing energy states. Using those we can reconstruct the propagator with a much smaller noise at large euclidean time distances.
In this work we present an efficient way to measure two-pion propagators and our results on reconstruction of the vector meson propagator with staggered fermions in a $(6^3\times 8.5)\,\textrm{fm}^4$ box.
}

\FullConference{The 41st International Symposium on Lattice Field Theory (LATTICE2024)\\
 28 July - 3 August 2024\\
Liverpool, UK\\}

\begin{document}
\maketitle
\section{Introduction}
The ongoing investigation of the anomalous magnetic moment of the muon, particularly through the $g-2$ experiment~\citep{Muong-2:2023cdq} and corresponding theoretical predictions~\citep{Aoyama:2020ynm}, has highlighted a significant discrepancy exceeding $4\sigma$. Lattice computations, on the other hand, tend to confirm the experimental results up to $0.9\sigma$~\citep{Borsanyi:2020mff,Boccaletti:2024guq,Djukanovic:2024cmq,Risch:2025}. A particularly intriguing aspect of these studies is the long-distance contribution that is dominated by the two-pion states that have a energy lower than the mass of the vector meson. However, the challenge of poor signal-to-noise ratios at these long distances complicates the analysis.

Various approaches have been proposed to address this issue. These include the bounding method \citep{Borsanyi:2020mff,Bazavov:2024eou}, the fitting method \citep{Lahert:2024vvu,Erben:2019nmx,Bruno:2019nzm,Blum:2024drk,Djukanovic:2024cmq}, and data-driven techniques~\citep{Boccaletti:2024guq}. This article focuses on the application of the fitting method using staggered fermions, as demonstrated in recent studies~\citep{Lahert:2021xxu,Frech:2023bzy,Lahert:2024vvu}. Given the high computational cost of connected two-pion diagrams, we present an efficient approach leveraging low-mode averaging to improve the signal quality, offering a promising pathway for more accurate results.
\section{$\pi\pi$ contributions to the vector correlator}
The study of taste-singlet vector mesons at rest within the staggered symmetry group framework, as detailed in various works~\citep{Lahert:2021xxu,Chreim:2024,Frech:2023bzy,Kilcup:1986dg,Golterman:1984dn,Golterman:1986jf,Golterman:2024xos,Baake:1981qe,Mandula:1983ut}, offers critical insights into the computation of two-pion contributions to the vector meson propagator. By employing Clebsch-Gordan coefficients~\citep{Sakata:1974hd} specific to the staggered symmetry group, it becomes possible to accurately determine these contributions. The methods for computing these coefficients, which are fundamental to the construction of the vector meson from two-pion states, are comprehensively discussed in~\citep{Frech:2023bzy}.

Considering a lattice of approximately $L = 6\,\mathrm{fm}$ in spatial extension with a spacing of around $a = 0.13\,\mathrm{fm}$, corresponding to a taste-breaking effect of $\Delta_{KS} = 39484\,\mathrm{MeV}^2$ as noted in~\citep{Boccaletti:2024guq}, this reveals that several two-pion states possess lower energies than the vector meson. These states are highlighted in red in table~\ref{tab:multi}.
\begin{table}[h]
    \centering
    \begin{tabular}{c||c|c|c|c|c}
                           &$\Vert \vec p \Vert^2 = 0$&$\Vert \vec p \Vert^2 = 1$&$\Vert \vec p \Vert^2 = 2$& $\Vert \vec p \Vert^2 = 3$& $\Vert \vec p \Vert^2 = 4$    \\
                           \hline\hline
         $\Vert \vec\xi \Vert^2 = 0$& 0&\textcolor{red}{1}&\textcolor{red}{1}&\textcolor{red}{1}&1\\
         $\Vert \vec\xi \Vert^2 = 1$& 0&\textcolor{red}{2}&\textcolor{red}{3}&2&2\\
         $\Vert \vec\xi \Vert^2 = 2$& 0&\textcolor{red}{2}&3&2&2\\
         $\Vert \vec\xi \Vert^2 = 3$& 0&1&1&1&1\\
    \end{tabular}
    \caption{Multiplicities of the the diagram up to momenta of $\Vert p \Vert^2 = 4$ in units of $\frac{2\pi}{L}$. For pions at rest there is no overlap to a vector-like state.}
    \label{tab:multi}
\end{table}

Due to the multiplicity of some of these spin-taste combination, a total of 10 two-pion states contribute to the vector meson propagator. The detailed Clebsch-Gordan coefficients for these contributions, which are crucial for the accurate construction of the vector meson from these two-pion states, are provided in the table below.
\begin{table}
\centering
\begin{tabular}{|c||c|c|}
\hline
&$\vec \xi^2 = 0(3)$&$\vec \xi^2 = 1(2)$\\
\hline\hline
$\vec p^2 = 1(4)$&$C^\alpha(\lambda \vec e_i,\vec 0) = \lambda \frac{1}{\sqrt{2}}\delta^{\alpha i}$&$C^\alpha(\lambda \vec e_i,\vec f_j) = \lambda \frac{1}{\sqrt{2}} \delta^{ij}\delta^{\alpha i}$\\
&&$C^\alpha(\lambda \vec e_i,\vec f_j) =  \frac{\lambda}{2}(1-\delta_{ij})\delta^{\alpha i}$\\
\hline
$\vec p^2 = 2$&$C^\alpha(\lambda \vec e_i + \mu e_j,\vec 0) =  \frac{\lambda}{2\sqrt{2}}\delta^{\alpha i}$&$C^\alpha(\lambda \vec e_i + \mu \vec e_j,\vec f_k) =  \frac{\lambda}{2\sqrt{2}}\vert\epsilon^{ijk}\vert\delta^{\alpha i}$ \\
&&$C^\alpha(\lambda \vec e_i + \mu \vec e_j,\vec f_k) = \frac{\lambda}{2\sqrt{2}}\delta^{ki}\delta^{\alpha i}$ \\
&&$C^\alpha(\lambda \vec e_i + \mu \vec e_j,\vec f_k) = \frac{\lambda}{2\sqrt{2}}\delta^{kj}\delta^{\alpha i}$ \\
\hline
$\vec p^2 = 3$&$C^\alpha(\lambda \vec e_i + \mu \vec e_j + \nu \vec e_k,\vec 0) =  \frac{\lambda}{2\sqrt{2}}\delta^{\alpha i}$&$C^\alpha(\lambda \vec e_i + \mu \vec e_j + \nu \vec e_k,\vec f_l) =  \frac{\lambda}{2\sqrt{2}}\delta^{il}\delta^{\alpha i}$ \\
&&$C^\alpha(\lambda \vec e_i + \mu \vec e_j + \nu \vec e_k,\vec f_l) =  \frac{\lambda}{4}\vert\epsilon^{jkl}\vert\delta^{\alpha i}$ \\
\hline
\end{tabular}
\caption{The Clebsch-Gordan coefficients needed for construcing the vector meson out of two-pion states. The computation of these coefficients as well as further explanations are given in~\citep{Frech:2023bzy}. Confirmations can be found in~\citep{Lahert:2024vvu,Lahert:2023ore}.}
\label{tab:CG}
\end{table}
\section{Computation of the connected $\pi\pi \to \pi\pi$ diagram}
\begin{figure}[h]
\begin{center}
\begin{tikzpicture}[very thick,q0/.style={->,DarkBlue,semithick,yshift=5pt,shorten >=5pt,shorten <=5pt}]

  % Loop
  \def\radius{0.9}
 
  \draw[] (0,-\radius) -- (2*\radius,-\radius) ;
  \draw[] (0,-\radius) -- (0,\radius) ;
  \draw[] (2*\radius,-\radius) -- (2*\radius,\radius) ;
  \draw[] (0,\radius) -- (2*\radius,\radius) ;
    \node[left] (1) at (0,0) {$u$};
        \node[right] (1) at (2*\radius,0) {$u$};
    \node[above] (1) at (\radius,\radius) {$\bar d$};
    \node[below] (1) at (\radius,-\radius) {$\bar d$};
    \node[above] (1) at (-0.2*\radius,\radius) {$\pi^{\dagger}(\vec x,0)$};
    \node[below] (1) at (-0.2*\radius,-\radius) {$\pi(\vec y,0)$};
    \node[above] (1) at (2.2*\radius,\radius) {$\pi(\vec z,t)$};
    \node[below] (1) at (2.2*\radius,-\radius) {$\pi^{\dagger}(\vec w,t)$};
    \node[below] (1) at (\radius,-1.8*\radius) {$a)$};
\def \offset{3.5}
\draw[] (\offset,-\radius) -- (\offset+2*\radius,-\radius) ;
  \draw[] (\offset,-0.9*\radius) -- (\offset+2*\radius,-0.9*\radius) ;
  \draw[] (\offset,\radius) -- (\offset+2*\radius,\radius) ;
  \draw[] (\offset,0.9*\radius) -- (\offset+2*\radius,0.9*\radius) ;
    \node[below] (1) at (\offset+\radius,\radius) {$u$};
    \node[above] (1) at (\offset+\radius,\radius) {$\bar d$};
    \node[below] (1) at (\offset+\radius,-\radius) {$u$};
    \node[above] (1) at (\offset+\radius,-\radius) {$\bar d$};
    \node[above] (1) at (\offset-0.2*\radius,\radius) {$\pi^{\dagger}(\vec x,0)$};
    \node[below] (1) at (\offset-0.2*\radius,-\radius) {$\pi(\vec y,0)$};
    \node[above] (1) at (\offset+2.2*\radius,\radius) {$\pi(\vec z,t)$};
    \node[below] (1) at (\offset+2.2*\radius,-\radius) {$\pi^{\dagger}(\vec w,t)$};
        \node[below] (1) at (\offset + \radius,-1.8*\radius) {$b)$};

\begin{feynman}
    \vertex (a) at (\offset+0.6*\radius,\radius);
    \vertex (b) at (\offset+1.7*\radius,-\radius);
    \diagram*{
      (a) -- [gluon] (b);
    };
\end{feynman}

    \draw[] (2*\offset,-\radius) -- (2*\offset,+\radius) ;
  \draw[] (2*\offset + 0.1*\radius,-\radius) -- (2*\offset + 0.1*\radius, +\radius) ;
 \draw[] (2*\offset+2*\radius,-\radius) -- (2*\offset+2*\radius,+\radius) ;
  \draw[] (2*\offset + 1.9*\radius,-\radius) -- (2*\offset + 1.9*\radius, +\radius) ;
    \node[left] (1) at (2*\offset,0) {$u$};
    \node[right] (1) at (2*\offset,0) {$\bar d$};
    \node[left] (1) at (2*\offset+2*\radius,0) {$u$};
    \node[right] (1) at (2*\offset+2*\radius,-0) {$\bar d$};
 \begin{feynman}
    \vertex (a) at (2*\offset,0.6*\radius);
    \vertex (b) at (2*\offset+2*\radius,-0.6*\radius);
    \diagram*{
      (a) -- [gluon] (b);
    };
    \end{feynman}
    
    \node[above] (1) at (2*\offset-0.2*\radius,\radius) {$\pi^{\dagger}(\vec x,0)$};
    \node[below] (1) at (2*\offset-0.2*\radius,-\radius) {$\pi(\vec y,0)$};
    \node[above] (1) at (2*\offset+2.2*\radius,\radius) {$\pi(\vec z,t)$};
    \node[below] (1) at (2*\offset+2.2*\radius,-\radius) {$\pi^{\dagger}(\vec w,t)$};
            \node[below] (1) at (2*\offset + \radius,-1.8*\radius) {$c)$};

\end{tikzpicture}

\begin{tikzpicture}[very thick,q0/.style={->,DarkBlue,semithick,yshift=5pt,shorten >=5pt,shorten <=5pt}]

  % Loop
  \def\radius{0.9}
 
  \draw[] (0,0) -- (2*\radius,-\radius) ;
  \draw[] (2*\radius,-\radius) -- (2*\radius,\radius) ;
  \draw[] (0,0) -- (2*\radius,\radius) ;
    \node[right] (1) at (2*\radius,0) {$\bar d$};
    \node[above] (1) at (\radius,0.5*\radius) {$u$};
    \node[below] (1) at (\radius,-0.5*\radius) {$u$};
    \node[left] (1) at (0,0) {$\rho^{\dagger}(\vec x,0)$};
    \node[above] (1) at (2.2*\radius,\radius) {$\pi(\vec z,t)$};
    \node[below] (1) at (2.2*\radius,-\radius) {$\pi^{\dagger}(\vec w,t)$};
            \node[below] (1) at (\radius,-1.8*\radius) {$d)$};
   \def\offset{5*\radius}
 \draw[] (\offset,0) -- (\offset + 2*\radius,0) ;
  \draw[] (\offset,0.1) -- (\offset + 2*\radius,0.1) ;
  \draw[] (0,0) -- (2*\radius,\radius) ;
    \node[above] (1) at (\offset +\radius,0.2*\radius) {$u$};
    \node[below] (1) at (\offset +\radius,-0.2*\radius) {$\bar d$};
    \node[left] (1) at (\offset +0,0) {$\rho^{\dagger}(\vec x,0)$};
    \node[left] (1) at (\offset +3.5*\radius,0) {$\rho(\vec z,0)$};
        \node[below] (1) at (\offset + \radius,-1.8*\radius) {$e)$};

\end{tikzpicture}
\end{center}
\caption{The diagrams that have to be implemented for GEVP. Diagram c) vanishes identically in the $I = 1$ case.}
\label{fig:diagrams}
\end{figure}
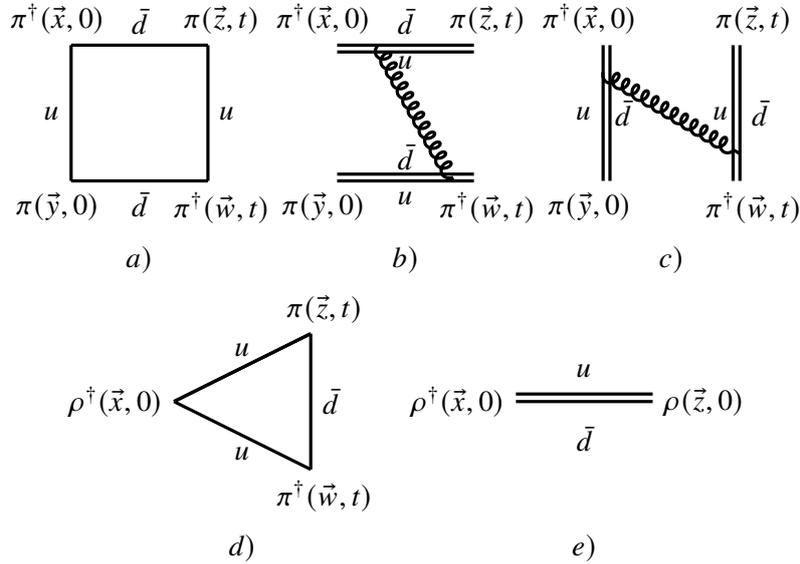
The general strategy to compute the diagrams that are shown in Figure~\ref{fig:diagrams} is to include wall sources $\zeta$ at an euclidean time $t_0$ and apply the operators of the spin-taste-momentum combination ($\rho/\pi_{\xi}(\vec p)$) as well as the inverse staggered operator ($M^{-1}_{t_1,t_2}$). One can easily convince oneself that diagram $c)$ vanishes identically for $I = 1$. While diagram $e)$ can be computed straightforwardly and diagram $b)$ is given by the product of two diagrams of this type, the two remaining diagrams require methods like sequential inversions in this general approach. For computing diagram $d)$, we can interchange the $\rho$ state and the two-pion state due to the time-reversibility of QCD. The corresponding scalar product is therefore given by
\begin{align}
C_{\rho\pi\pi}(t) = 
\langle\zeta_{t_0}M^{-1}_{t_0,t_0 + t}\rho  \vert M^{-1}_{t+t_0,t_0} \pi M^{-1}_{t_0,t_0}\pi \zeta_{t_0}\rangle.
\end{align}
Inversions only have to be applied at $t_0$ time slices, and therefore the number of inversions does not scale with the time-extent of the lattice or the domain of $C_{\rho\pi\pi}$. Nevertheless, this is not the case for the connected part of the two-pion propagator, because it requires two operators at each time slice:
\begin{align}
\label{eq:4pion}
C_{\pi\pi\pi\pi}^{\textit{conn.}}(t) = 
\langle\zeta_{t_0}M^{-1}_{t_0,t_0 + t}\pi \color{red}M^{-1}_{t_0+t,t_0 + t}\color{black}\pi \vert M^{-1}_{t+t_0,t_0} \pi M^{-1}_{t_0,t_0}\pi \zeta_{t_0}\rangle.
\end{align}
The red inversion’s column and row have a time dependence. So overall, we have a number of inversions proportional to
\begin{align}
N_\zeta \times N_O \times T/a
\end{align}
where $T/a$ is the time extent in lattice units, $N_\zeta$ is the number of random sources, and $N_O$ is the number of pion operators (which has a strong volume and spacing dependence, typically ranging between 10 and 100).

We aim to reduce the cost by splitting up $M^{-1}$ into its eigenvalue and residual parts. The lowest eigenvalues $\lambda$ with the corresponding eigenvectors are labeled by $i$, the projection of $M^{-1}$ on the orthogonal complement of the lowest eigenvectors is called $M_r^{-1}$:
\begin{align}
M^{-1} &= \sum_i \frac{1}{\lambda_i}\ket{i}\bra{i} + M_r^{-1}\\
&\approx\sum_i \frac{1}{\lambda_i}\ket{i}\bra{i} + \sum_\sigma \ket{\sigma}\bra{\sigma} M_r^{-1}
\end{align}
Eigenvectors and values are already computed from the preconditioning of the Dirac operator. The vectors $\sigma$ are stochastic sources normalised to give an approxiamtion of the identity. Now the number of inversions is dominated by the ket-vectors in Equation~\ref{eq:4pion}, so we removed the $T/a$ factor in the total amount of inversions.

\begin{figure}
\centering
\includegraphics[width =  \textwidth]{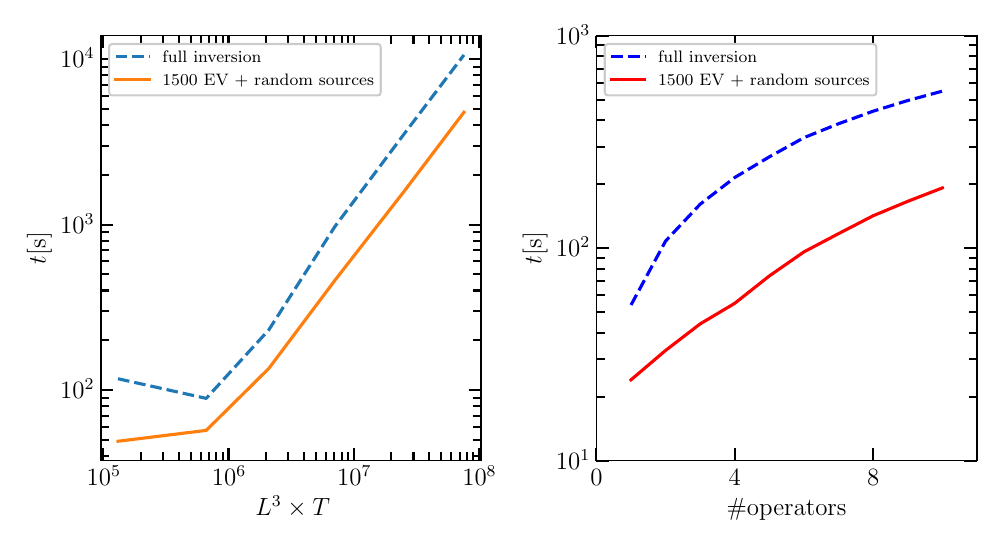}
\caption{On the right-hand side, one can see the volume scaling of the LMA method compared to the usual sequential inversion. On the right-hand side, one can see the scaling with the number of operators. Both plots use 1500 eigenvectors and 18 random wall sources $\zeta$. To estimate the residual part, we include $T/2$ random sources $\sigma$.}
\label{fig:scaling}
\end{figure}

In figure \ref{fig:scaling}, one can see the scaling of the computational time of the LMA method compared to the usual sequential inversion with respect to the volume and the number of operators. One can observe a better prefactor in the overall scaling, as well as a slightly better volume scaling with the new method.

\begin{figure}
\centering
\includegraphics[width = \textwidth]{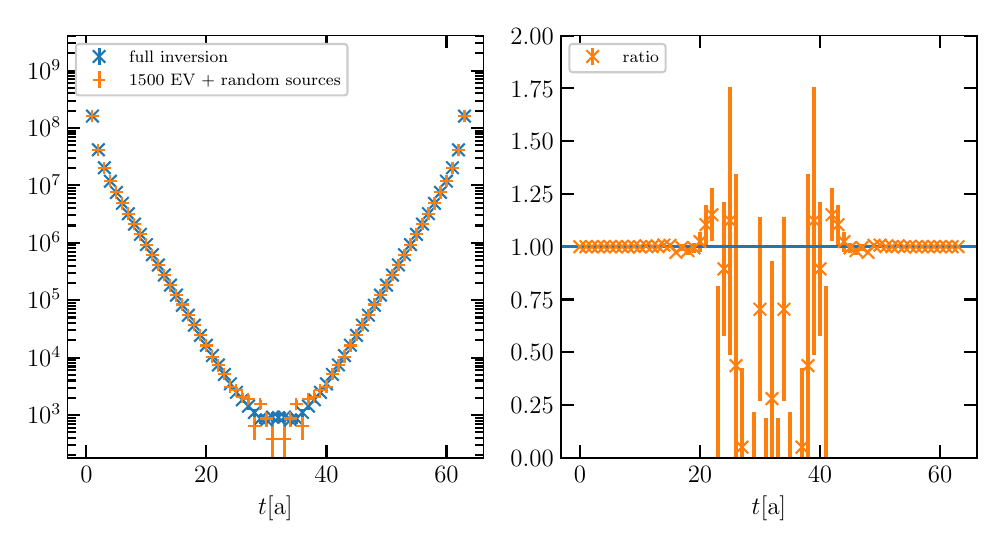}
\caption{Two Goldstone pions with momenta $\vec p_1 = (0,1,1)$ and $\vec p_2 = (0,-1,-1)$. The full propagators are on the right-hand side, while their ratio is shown on the left-hand side. We use 18 random wall sources for both methods. For the LMA method, we use 1500 eigenvectors and 32 sources to estimate the residual part.}
\label{fig:comp}
\end{figure}

In figure \ref{fig:comp}, we compare the two-pion propagators with Goldstone taste and $\vert \vec p \vert^2 = 2$. One can see that the new method has larger statistical noise, especially at large distances. However, this is not problematic, as we aim to extract the contributing mode using a GEVP from the earlier time slices. The comparison presented in figure 3 is representative of the average difference oberved across the tastes and momenta considered. However, smaller momenta and local tastes see less difference betwenn the two methods, and larger momenta and non-local tastes see more difference.

\section{Reconstruction of the vector meson propagator}
As mentioned in~\citep{Frech:2023bzy}, the two-pion states are constructed in such a way that they share all quantum numbers with the $\rho$-meson, except for the energy. More generally, the $\rho$-propagator, as well as the $\pi\pi$-propagator, consists of different linear combinations of states with certain quantum numbers and varying energies. To obtain the energy eigenstates, we solve the generalized eigenvalue problem (GEVP):
\begin{align}
   \mathbf{C}(t_0 + \mathrm{d}t)V(t_0 + \mathrm{d}t,t_0) = \lambda(t_0 + \mathrm{d}t,t_0)\mathbf{C}(t_0)V(t_0 + \mathrm{d}t,t_0),
\end{align}
where the correlation matrix is
\begin{align}
   \mathbf{C}(t) = \begin{pmatrix}
    C_{\rho\rho}(t) & C_{\rho\pi\pi}(t) \\
    C_{\pi\pi\rho}(t) & C_{\pi\pi\pi\pi}(t)
    \end{pmatrix}.
\end{align}
Here, $t$ and $t_0$ are fixed values. The single-state correlation functions are then given by
\begin{align}
    \tilde{\lambda}_i(t) = V_i^\dagger C(t) V_i,
\end{align}
where there are more than one two-pion state (in this case, ten). 

\begin{figure}
    \includegraphics[width=0.48\textwidth]{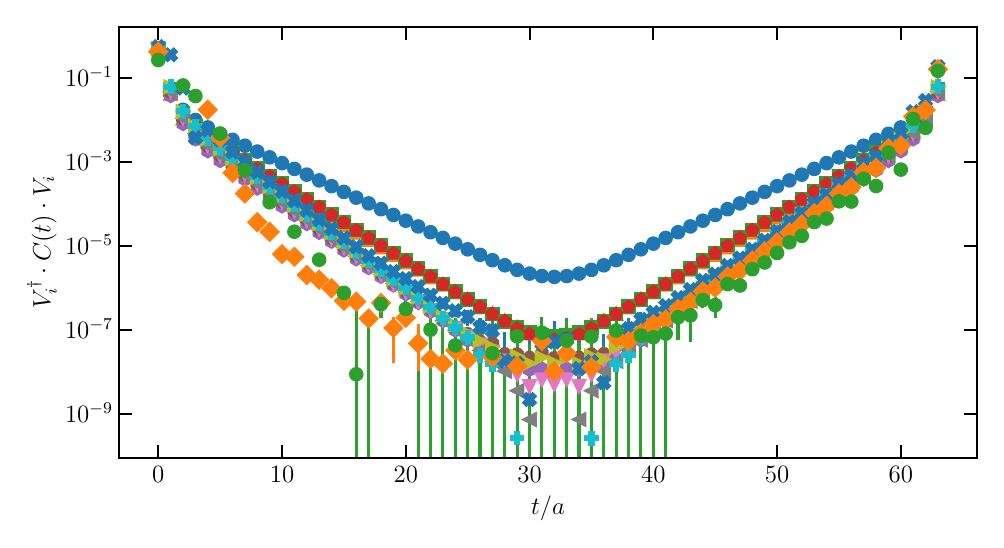}
    \includegraphics[width=0.48\textwidth]{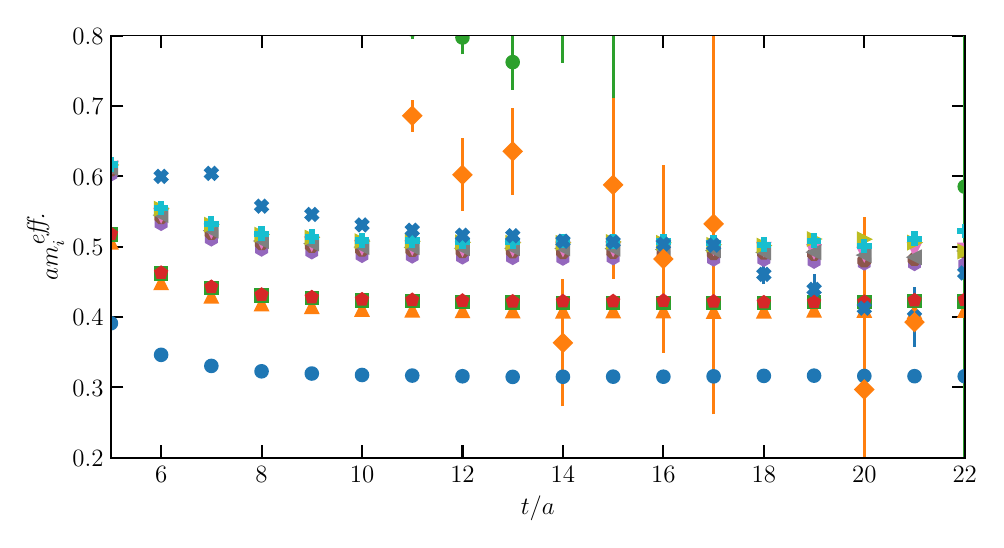}
    \caption{Right hand side: Eigenvalues $\tilde \lambda(t)$ of the GEVP including all the $\pi\pi$-propagators with $2(p_i^2  + m_{\pi_\xi}^2) \leq m_\rho^2$  as well as the (shifted) $\rho$-propagators. RIght hand side: Effective energies  $M^{\textit{eff}}_i(t)$ of the states shown on the left hand side.}
    \label{fig:states}
\end{figure}

To remove higher-state contaminations, we use the pencil-of-functions method~\citep{DeTar:2014gla}, applying different shifts to the $\rho$-propagator:
\begin{align*}
   \mathbf{C}(t)= \begin{pmatrix}
    C_{\rho\rho}(t-4) & C_{\rho\rho}(t-3) & C_{\rho\rho}(t-2) & C_{\rho\pi\pi}(t-2) \\
    C_{\rho\rho}(t-3) & C_{\rho\rho}(t-2) & C_{\rho\rho}(t-1) & C_{\rho\pi\pi}(t-1) \\
    C_{\rho\rho}(t-2) & C_{\rho\rho}(t-1) & C_{\rho\rho}(t)   & C_{\rho\pi\pi}(t)   \\
    C_{\pi\pi\rho}(t-2) & C_{\pi\pi\rho}(t-1) & C_{\pi\pi\rho}(t)   & C_{\pi\pi\pi\pi}(t)
    \end{pmatrix},
\end{align*}
resulting in a 13-dimensional\footnote{10 $\pi\pi$-propagators (see Table \ref{tab:multi}) and 3 $\rho$-propagators} GEVP in total. We choose $t_0 = 5a$ and $\mathrm{d}t = a$. The eigenvalues $\tilde{\lambda}_i$ are shown on the left-hand side of figure \ref{fig:states}. Two noisy states, which correspond to higher excitations, are filtered out using the pencil-of-functions method. 

We fit energy plateaus to the effective energies:
\begin{align}
    M^{\textit{eff}}_i(t) = \frac{1}{\Delta}\cosh^{-1}\left(\frac{\tilde{\lambda}_i(t+\Delta) + \tilde{\lambda}_i(t-\Delta)}{2\tilde{\lambda}_i(t)}\right),
\end{align}
where $\Delta = 2a$. The effective energies are displayed on the right-hand side of figure~\ref{fig:states}.

\begin{figure}
    \includegraphics[width=0.48\textwidth]{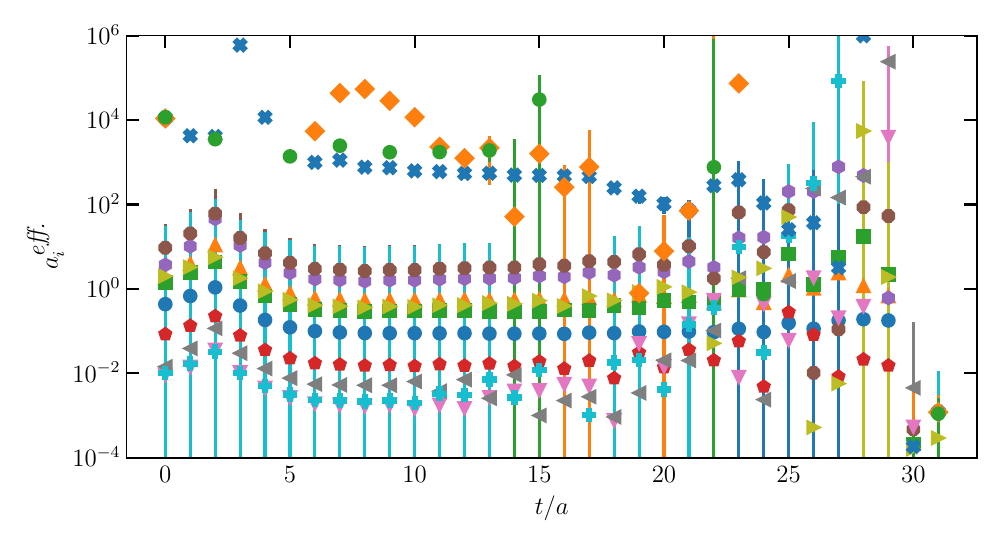}
    \includegraphics[width=0.48\textwidth]{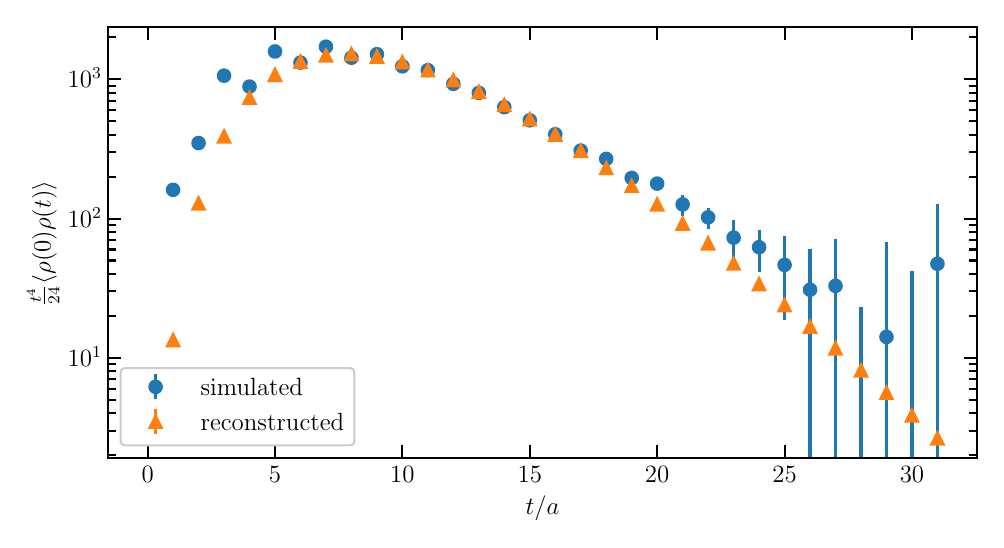}
    \caption{Left hand side: Coefficents $a_i^{\textit{eff.}}(t)$ of the considered states. Right hand: Reconstruction of the vector meson propagator $\rho_{\textrm{rec.}}(t) $compared to the simulated propagator $\rho(t)$.}
    \label{fig:recosntruction}
\end{figure}

The decomposition of the $\rho$ propagator into different energy states is given by:
\begin{align*}
    \rho_{\textrm{rec.}}(t) = \sum_i a_i \exp\left(-M_i t\right),
\end{align*}
where the coefficients can be computed using
\begin{align*}
    a_i^{\textit{eff.}}(t) = \frac{\left(v_i^T \cdot \mathbf{C}_{\rho\,\bullet}(t)\right)^2}{v_i^T \cdot \mathbf{C}(t) \cdot v_i} \exp\left(M^{\textit{eff.}}_i(t) \cdot t\right),
\end{align*}
where $\mathbf{C}_{\rho\,\bullet}(t)$ denotes the column of $\mathbf{C}(t)$ that correspond to the unshifted vector propagator.
Subsequently, we fit plateaus to the effective coefficients, as shown on the left-hand side of figure~\ref{fig:recosntruction}. Although the errors are large, most of the contribution comes from a single state and we can reconstruct the $\rho$ propagator using the available data. The first moment of this reconstruction is shown on the right-hand side of figure~\ref{fig:recosntruction}.

\section{Conclusion}
In this work, we presented a refined approach to computing the connected $\pi\pi \to \pi\pi$ diagram within the framework of staggered fermions, specifically focusing on the use of low-mode averaging (LMA) to enhance signal quality. By reducing the computational cost associated with the inversion of the Dirac operator, we significantly improved the efficiency of the calculation, making it more feasible to study long-distance contributions to the vector meson propagator. The LMA method provides a better scaling behavior compared to traditional sequential inversions, especially in terms of volume and operator count, as demonstrated in Figs.~\ref{fig:scaling} and~\ref{fig:comp}.

Despite some loss in signal quality at large momenta, the GEVP approach allows us to reliably extract relevant physics from the early time slices, where the signal remains strong. Future work will focus on further optimization of this method, potentially incorporating smearing of the vector operators to enhance the overlap of vector meson and two-pion propagators. This study lays the foundation for more precise lattice calculations of the anomalous magnetic moment of the muon.

\acknowledgments
The computations were performed on HAWK at the High Performance Computing Center in Stuttgart, SuperMUC at Leibniz-Rechenzentrum, Munich and JUWELS at Juelich Supercomputing Centre. We thank the Gauss Centre for Supercomputing, PRACE and GENCI (grant 52275) for awarding us computer time on these machines.

\end{document}